\documentclass[aps,prl,onecolumn,showpacs,amsmath,amssymb,floatfix,footinbib,superscriptaddress]{revtex4-1}
\usepackage{amsmath,natbib,graphicx,subfigure,amssymb,graphics,amsmath,mathrsfs,CJK,color}
\usepackage{multirow,fancyhdr,color,bm,tabularx,psfrag,dcolumn}
\usepackage[colorlinks=true,linkcolor=blue,urlcolor=blue,citecolor=blue]{hyperref}
\usepackage{tikz,xcolor,hyperref}% Make Orcid icon
\definecolor{lime}{HTML}{A6CE39}
\DeclareRobustCommand{\orcidicon}{%
    \begin{tikzpicture}
    \draw[lime, fill=lime] (0,0)
    circle [radius=0.16]
    node[white] {{\fontfamily{qag}\selectfont \tiny ID}};\draw[white, fill=white] (-0.0625,0.095)
    circle [radius=0.007];
    \end{tikzpicture}
    \hspace{-2mm}}
\foreach \x in {A, ..., Z}
{\expandafter\xdef\csname orcid\x\endcsname{\noexpand\href{https://orcid.org/\csname orcidauthor\x\endcsname}{\noexpand\orcidicon}}}

\usepackage[left]{lineno}%[switch]
\begin{document}

\title{Effect of spontaneously generated coherence on left-handedness in a degeneracy atomic system}
%\thanks{}

\author{Shun-Cai Zhao\orcidA{}}
\email[Corresponding author: ]{zscnum1@126.com.}
\affiliation{Physics department , Kunming University of Science and Technology, Kunming, 650093, PR China}

%\date{\today}

\begin{abstract}
A theoretical investigation is carried out into the effect of
spontaneously generated coherence(SGC) on the left-handedness in a
four-level Y-type atomic system with two highest nearly degenerate
lying levels. It is found, with the spontaneously generated
coherence intensity enhancing, the atomic system gradually displays
left-handedness with simultaneous negative permittivity and
permeability. And the refractive index enhances with the increasing
intensity of SGC.However, the absorption is suppressed by the SGC
effect when the SGC has a large intensity.When the probe field is
near-resonant coupled to the atomic system, the appearance of SGC
doesn't always change the permeability from positive to negative and
allow for left-handed behavior,unless the SGC reaches a large
intensity.
\begin{description}
\item[PACs]{42.50.Gy}
\item[Keywords]{spontaneously generated coherence, , left-handedness, negative permittivity and permeability}
\end{description}
\end{abstract}

\maketitle
\section{Introduction}

Quantum interference in quantum optics occurs when there are two or
more identical channels in a transition. The simplest example is
classic Fano interference [1-2] from a continuum of upper states to
a ground state. Another intriguing quantum effect is the
interference of two decay channels with nonorthogonal
electric-dipole transition matrix elements. This is named as
¡°spontaneously generated coherence¡±(SGC)[3-4] which was first
suggested by Agarwal [4] who showed that the spontaneous emission
from a degenerate V-type three-level atom is sensitive to the mutual
orientation of the atomic dipole moments. If they are parallel a
suppression of spontaneous emission can appear and a part of the
population can be trapped in the excited levels. SGC is one of the
two mechanisms of generating quantum coherence applied by incoherent
processes,i.e., spontaneous emission [5], and the other one is by
coherent fields, such as laser fields [6-9] or microwave fields
[10]. Here we are particularly interested in the former case and
impressive efforts have been made to investigate it in the last few
decades[11-20]. SGC requires two close-lying levels be nearly
degenerate and the atomic dipole moments be nonorthogonal when the
atom is placed in free space. And it is responsible for many
important physical phenomena, which are potentially applied in
lasing without population inversion [11], coherent population
trapping (CPT) [12], group velocity reduction [13], ultrafast
all-optical switching [14], transparent high-index materials [21],
high-precision spectroscopy and magnetometer [22], modified quantum
beats [23], dark-state polaritons [24], quantum information and
computing [25], etc. In order to observe the phenomena based on SGC,
a few methods have been proposed to simulate this intriguing effect.
SGC can be simulated by a dc field [26], a microwave field [27], or
a laser field [28-30].Very recently, there is experimental evidence
that SGC plays a role in charged GaAs quantum dots [31], which have
been proposed as elements in quantum-information networks. All the
interesting features due to SGC could have useful applications in
laser physics and other areas of quantum optics.

Although some works[32-33] for the realization of negative
refraction have been done, the dependence of SGC effect on the
left-handedness has never been investigated to our best knowledge.
Materials with left-handedness (medium having negative permittivity
and permeability simultaneously) which originally suggested by
Veselago [34],promise many surprising and even counterintuitive
electromagnetical and optical effects such as the reversals of both
Doppler shift and Cerenkov radiation [34], negative
Goos-H$\ddot{a}$anchen shift [35], amplification of evanescent waves
[36], sub-wavelength focusing [37] and so on. Up to now, several
interesting approaches have been proposed to realize LHM, such as
the method of artificial composite metamaterials [38-40], the method
of specific photonic crystal structures [41-42], the method of
chiral materials [43,44], and the method of photonic resonant media
[45-48]. In this paper, we investigate the effects of SGC on
left-handedness in a four-level Y-type atomic system.When the two
highest-lying levels of atomic system are nearly degenerate and
coupled by the same vacuum radiation field to the intermediate
state, the system can create SGC due to the quantum interference
between the two spontaneous decay channels.

And Our paper is organized as follows. In Sec. II we describe the
model and present the density-matrix equations of motion for the
system. In Sec.III we discuss the effects of SGC on the
left-handedness. The dependence of simultaneous negative
permittivity and permeability with SGC is discussed here. In Sec.IV,
the conclusion is presented .

\section{Theoretical analysis}
\begin{center}
\begin{figure}[h!]
  \centering
  \includegraphics[width=2.5in]{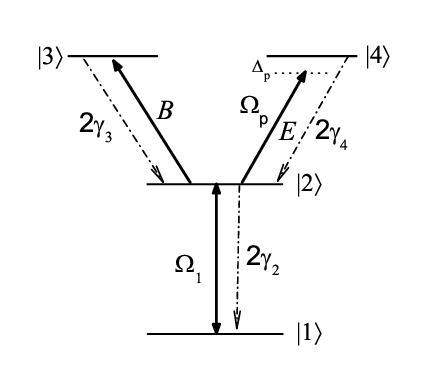}
  \hspace{0in}%
  \caption{{\bf Fig.1.} Schematic representation of the relevant atomic energy levels.}
\hspace{0in}%
\end{figure}\label{Fig.1}
\end{center}
We consider a four-level Y-type atomic system as shown in Fig.1. A
resonant coupling field $\Omega_{1}$ drives the transition between
levels $|1\rangle$ and$|2\rangle$ while a probe field $\Omega_{p}$
is applied to the transition $|2\rangle$ and $|4\rangle$. Its
electric (E) and magnetic (B) components of the probe field couple
state$|2\rangle$ to state $|4\rangle$ by an electric dipole
transition, and to state $|3\rangle$ by a magnetic dipole
transition. The parities in this system are $|2\rangle$ even,
$|3\rangle$ even, $|4\rangle$ odd or vice versa.
 2$\gamma_{3}$ and 2$\gamma_{4}$ are the spontaneous decay rates from levels $|3\rangle$ and $|4\rangle$ to
level $|2\rangle$, respectively, and 2$\gamma_{2}$ corresponds to
the decay rate from $|2\rangle$ to$|1\rangle$.In the interaction
picture the density-matrix equations of motion in the rotating-wave
approximations can be written as

\begin{equation}
\dot{\rho_{11}}=2\gamma_{2}\rho_{22}+i\Omega_{{1}}(\rho_{21}-\rho_{12})
\end{equation}
\begin{equation}
\dot{\rho_{33}}=2\gamma_{3}\rho_{33}-p\sqrt{\gamma_{3}\gamma_{4}}(\rho_{34}+\rho_{43})
\end{equation}
\begin{equation}
\dot{\rho_{44}}=-2\gamma_{4}\rho_{44}-p\sqrt{\gamma_{3}\gamma_{4}}(\rho_{34}+\rho_{43})+i\Omega_{{p}}(\rho_{24}-\rho_{42})
\end{equation}
\begin{equation}
\dot{\rho_{12}}=-\gamma_{2}\rho_{12}+i\Omega_{{1}}(\rho_{22}-\rho_{11})-i\Omega_{{p}}\rho_{14}
\end{equation}
\begin{equation}
\dot{\rho_{13}}=-\gamma_{3}\rho_{13}+i\Omega_{{1}}\rho_{23}-p\sqrt{\gamma_{3}\gamma_{4}}\rho_{14}
\end{equation}
\begin{equation}
\dot{\rho_{14}}=-(\gamma_{4}-i\Delta_{p})\rho_{14}-p\sqrt{\gamma_{3}\gamma_{4}}\rho_{13}+i\Omega_{1}\rho_{24}-i\Omega_{{p}}\rho_{12}
\end{equation}
\begin{equation}
\dot{\rho_{23}}=-(\gamma_{2}+\gamma_{3})\rho_{23}+i\Omega_{{1}}\rho_{13}+i\Omega_{{p}}\rho_{43}-p\sqrt{\gamma_{3}\gamma_{4}}\rho_{24}
\end{equation}
\begin{equation}
\dot{\rho_{24}}=-[\gamma_{2}+\gamma_{4}-i\Delta_{p}]\rho_{24}+i\Omega_{p}(\rho_{44}-\rho_{22})+i\Omega_{1}\rho_{14}-p\sqrt{\gamma_{3}\gamma_{4}}\rho_{23}
\end{equation}
\begin{equation}
\dot{\rho_{34}}=-(\gamma_{3}+\gamma_{4}-i\Delta_{p})\rho_{34}-i\Omega_{p}\rho_{32}-p\sqrt{\gamma_{3}\gamma_{4}}(\rho_{33}+\rho_{44})
\end{equation}

The above equations are constrained by
$\rho_{11}+\rho_{22}+\rho_{33}+\rho_{44}=1$ and
$\rho_{ij}=\rho_{ji}^{\ast}$.$\Delta_{p}=\omega_{24}-\omega_{p}$
means the detuning of the probe field from the optical transition.
The terms with $p\sqrt{\gamma_{3}\gamma_{4}}$ represent the quantum
interference resulting from the cross coupling between spontaneous
emission paths $|3\rangle$-$|2\rangle$ and$|4\rangle$-$|2\rangle$.

The parameter p denotes the alignment of the two dipole moments.If
the two dipole moments are orthogonal to each other, i.e., p=0, the
SGC effect disappears. While for the  parallel or antiparallel case
we have p=$\pm1$, the interference between the spontaneous emissions
and the SGC effect reach the maximum.So we can depict dynamic
intensity of the SGC effect by parameter p. With the restriction
that each field acts only on one transition, the Rabi frequencies
$\Omega_{1}$ and $\Omega_{P}$ are represented by
$\Omega_{1(P)}$=$\Omega_{1(P)}^{0}\sqrt{1-p^{2}}$. It should be
noted that only for small energy spacing between the two
highest-lying levels are significant; otherwise the oscillatory
terms will average out to zero and thereby the SGC effect
vanishes[49-51].

In the following, we will discuss the electric and magnetic
responses of the medium to the probe field. It should be noted that
here the atoms are assumed to be nearly stationary(e.g.,at a low
temperature)and hence any Doppler shift is neglected. When
discussing how the detailed properties of the atomic transitions
between the levels are related to the electric and magnetic
susceptibilities, one must make a distinction between macroscopic
fields and the microscopic local fields acting upon the atoms in the
vapor. In a dilute vapor, there is little difference between the
macroscopic fields and the local fields that act on any
atoms(molecules or group of molecules)[52]. But in dense media with
closely packed atoms(molecules),the polarization of neighboring
atoms(molecules) gives rise to an internal field at any given atom
in addition to the average macroscopic field, so that the total
fields at the atom are different from the macroscopic fields[52]. In
order to achieve the negative permittivity and permeability, here
the chosen vapor with atomic concentration $N=5\times10^{24}m^{-3}$
should be dense, so that one should consider the local field effect,
which results from the dipole-dipole interaction between neighboring
atoms. In what follows we first obtain the atomic electric and
magnetic polarization, and then consider the local field correction
to the electric and magnetic susceptibilities(and hence to the
permittivity and permeability)of the coherent vapor medium. With the
formula of the atomic electric polarizations
$\gamma_{e}=2d_{42}\rho_{24}/\epsilon_{0}E_{p}$,where
$E_{p}=\hbar\Omega_{p}/d_{42}$ one can arrive at
\begin{eqnarray}
\gamma_{e}=\frac{2d_{42}^2\rho_{24}}{\epsilon_{0}\hbar\Omega_{p}}\
\end{eqnarray}
In the similar fashion, by using the formulae of the atomic magnetic
polarizations $\gamma_{m}=2\mu_{0}\mu_{23}\rho_{32}/B_{p}$ [52], and
the relation of between the microscopic local electric and magnetic
fields $E_{p}/B_{p}=c$ we can obtain the explicit expression for the
atomic magnetic polarizability. Where $\mu_{0}$is the permeability
of vacuum, c is the speed of light in vacuum. Then, we have obtained
the microscopic physical quantities $\gamma_{e}$and$\gamma_{m}$.
Thus, the coherence $\rho_{32}$ drives a magnetic dipole, while the
coherence $\rho_{24}$ drives an electric dipole. However, what we
are interested in is the macroscopic physical quantities such as the
electric and magnetic susceptibilities which are the electric
permittivity and magnetic permeability. The electric and magnetic
Clausius-Mossotti relations can reveal the connection between the
macroscopic and microscopic quantities. According to the
Clausius-Mossotti relation [53], one can obtain the electric
susceptibility of the atomic vapor medium
\begin{eqnarray}
\chi_{e}=N\gamma_{e}\cdot{{{{(1-\frac{N\gamma_{e}}{3})}}}}^{-1}
\end{eqnarray}
The relative electric permittivity of the atomic medium reads
$\varepsilon_{r}=1+\chi_{e}$.In the meanwhile,the magnetic
Clausius-Mossotti [53]
\begin{eqnarray}
\gamma_{m}=\frac{1}{N}(\frac{\mu_{r}-1}{\frac{2}{3}+\frac{\mu_{r}}{3}})
\end{eqnarray}
shows the connection between the macroscopic magnetic permeability
$\mu_{r}$ and the microscopic magnetic polarizations $\gamma_{m}$.
It follows that the relative magnetic permeability of the atomic
vapor medium is
\begin{eqnarray}
\mu_{r}=\frac{1+\frac{2}{3}N\gamma_{m}}{1-\frac{1}{3}N\gamma_{m}}
\end{eqnarray}
Substituting the expressions of $\varepsilon_{r}$ and $\mu_{r}$ into
$n=-\sqrt{\varepsilon_{r}\mu_{r}}$ [34], we can get the refractive
index of left-handed materials. In the above, we obtained the
expressions for the electric permittivity,magnetic permeability and
refractive index of the four-level atomic medium. In the section
that follows, we will get solutions to the density-matrix equations
(1) under the stead-state condition.

\section{Results and discussions}

\begin{center}
\begin{figure}[h!]
  \centering
  \includegraphics[width=2.7in]{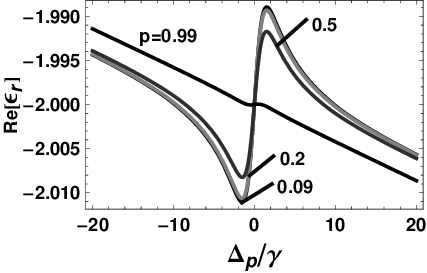}
  \hspace{0in}%
  \caption{{\bf Fig.2.} Real parts of the permittivity versus the rescaled
detuning parameter $\Delta_{p}$/$\gamma$ with: $\Omega_{1}^{0}$=
$10\gamma$, $\Omega_{p}^{0}$=0.2$\gamma$,
$\gamma_{2}$=$\gamma_{3}$=$\gamma_{4}$=0.8$\gamma$. The curves are
labeled by p values.}
\hspace{0in}%
\end{figure}\label{Fig. 2}

\begin{figure}[h!]
  \centering
  \includegraphics[width=2.5in]{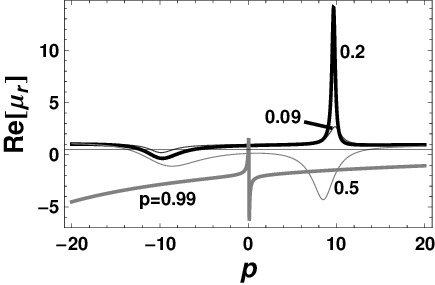}
  \hspace{0in}%
  \caption{{\bf Fig.3.} Real parts of the permeability versus the rescaled
detuning parameter $\Delta_{p}$/$\gamma$. The curves are labeled by
p values.}
\hspace{0in}%
\end{figure}\label{Fig. 3}
\end{center}

For simplicity, we scale all the parameters by $\gamma=10^{8}$,
$\Omega_{1}^{0}$=$10\gamma$, $\Omega_{p}^{0}$=0.2$\gamma$,
$\gamma_{2}$=$\gamma_{3}$=$\gamma_{4}$=0.8$\gamma$ during our
calculation. Analyzing the left-handed behavior in the atomic
system, the plus or minus for the real part of permittivity
$\epsilon_{r}$ and permeability $\mu_{r}$ is the content for
considered. We show in Fig.2 the real part of the relative electric
permittivity $\epsilon_{r}$ as a function of the probe rescaled
detuning $\Delta_{p}$/$\gamma$. Different curves corresponds to
different values of p noted in the caption of Fig.2. We noticed that
the real part of relative electric permittivity $\epsilon_{r}$
always remains negative in the range of detuning considered here
while varying the value of p, and the line shape is similar in
Fig.2. The contribution of local field due to dipole-dipole
interaction plays an important role in the real parts of
permittivity remaining negative values, because the atomic density
is assumed very high here.We also noticed that when p=0.99 which
means a strong interference between the two spontaneous emission
channels and the SGC effect is strongest, the fluctuation range
Re[$\epsilon_{r}$] is the smallest near the resonant point.

In Figs.3,the real part of magnetic permeability $\mu_{r}$ versus
$\Delta_{p}$/$\gamma$ is plotted with the same parameters as used in
Fig. 2. It shows that when p=0.09, Re[$\mu_{r}$] is always positive
in the detuning range. The atomic system shows no left-handed
behavior with the weak SGC effect caused by the interference between
the spontaneous emission channels. Because the left-handed material
(LHM), should possess negative real parts of both dielectric
permittivity $\epsilon_{r}$ and magnetic permeability $\mu_{r}$ over
the same frequency band. The magnetic response of our atomic system
is quite different when the SGC gets a large intensity.
Re[$\mu_{r}$] is firstly negative value in a small frequency band
near $\Delta_{p}=10\gamma$ when p reaches the value of 0.2 as shown
in Fig.2. Two larger frequency ranges show that Re[$\mu_{r}$] is
negative when p=0.5. Further increasing p to the value of 0.99, the
negative Re[$\mu_{r}$] is in all the detuning range except the
resonant point. The SGC is stronger, the wider and much more
frequency ranges for negative permeability. That is, the SGC effect
greatly enhances the magnetic response of a dense atomic gas so that
one can realize negative refractive index. Here, we stress that the
refractive index for left-handed materials is given by
$n=-\sqrt{\varepsilon_{r}\mu_{r}}$ where the symbol$``-"$corresponds
to the only case of both $\varepsilon_{r}$ and $\mu_{r}$ having
negative real parts.The enhancement of SGC results the
simultaneously negative real parts for permittivity and
permeability. Hence, we attribute the emergence of left-handedness
to the enhancement of the SGC effect which induced by the
interference between the two spontaneous emission channels from the
two highest-lying levels in the atomic system.

\begin{center}
\begin{figure}[h!]
  \centering
  \includegraphics[width=2.3in]{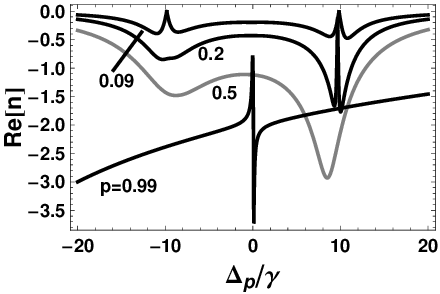}
  \hspace{0in}%
  \includegraphics[width=2.7in]{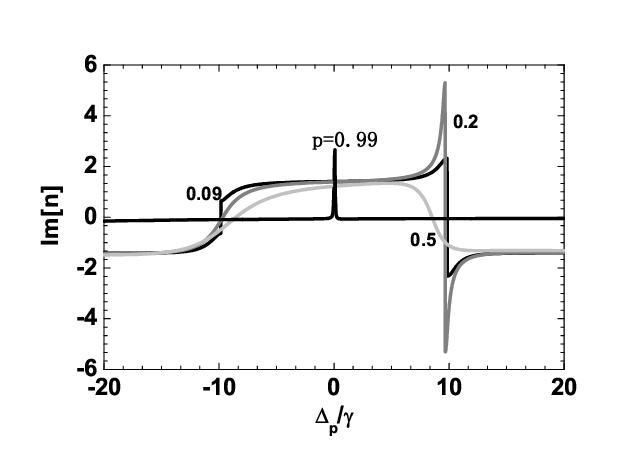}
  \hspace{0in}%
  \caption{{\bf Fig.4.} The refractive index versus the
rescaled detuning $\Delta_{p}$/$\gamma$.}
\hspace{0in}%
\end{figure}\label{Fig. 4}
\end{center}

In Fig.4, the refractive index is plotted for different values of
the parameter p with the same parameters as used in Fig.2. As
observed in Fig.4, the amplitude of the real parts of refractive
index gradually increases as the intensifying of SGC. The maximum
value of Re[n]are -1.7, -2.9, -3.7 when p is varied by 0.2, 0.5,
0.99, respectively. Obviously, the intensity of SGC effect
influences its peak value.The imaginary part of refractive index
which depicts the atomic system absorption property is plotted on
the right side of Fig.4.Two transparency windows turn into two
transparency ranges beside the resonance. The absorption is
suppressed by the SGC effect when it has a large intensity shown by
Fig.4.

Above, we discussed the left-handedness as functions of the detuning
parameter $\Delta_{p}$/$\gamma$. In order to get a deeper insight
into the effect of SGC on left-handedness, in Fig.5 and Fig.6, we
plot the case of the probe field near-resonantly coupling
$|2\rangle$ and $|4\rangle$ ($\Delta_{p}$=$10^{-16}$$\gamma$) with
other parameters being the same as used in Fig. 2. And here, the
characteristic parameters of left-handedness is depicted as the
functions of p which will value from  the orthogonality to alignment
of the two dipole moments, because the conclusion of anti-alignment
will be symmetric to the alignment.

\begin{center}
\begin{figure}[h!]
  \centering
  \includegraphics[width=2.4in]{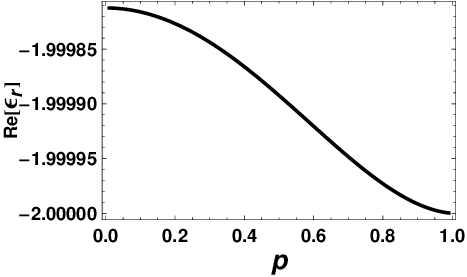}
  \includegraphics[width=2.4in]{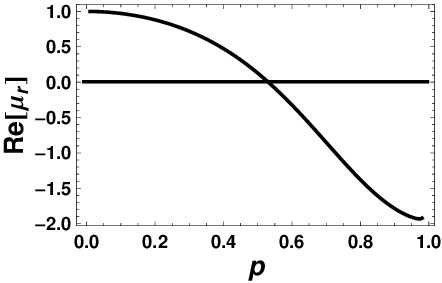}
  \hspace{0in}%
  \makeatletter\def\@captype{figure}\makeatother\caption{{\bf Fig.5.} Real parts of the permittivity and permeability versus p with the probe field
  near-resonant coupling $|2\rangle$-$|4\rangle$ .}
\hspace{0in}%
\end{figure}\label{Fig. 5}

\begin{figure}[h!]
  \centering
  \includegraphics[width=2.5in]{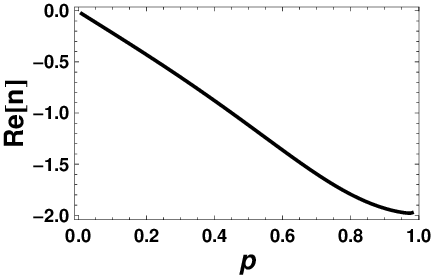}
  \includegraphics[width=2.5in]{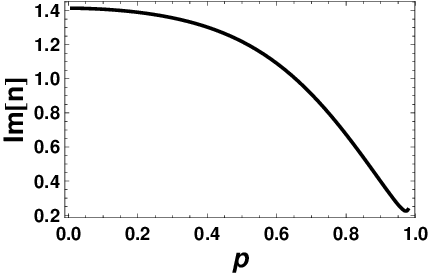}
  \hspace{0in}%
  \caption{{\bf Fig.6.} The refractive index verus p with the
probe field near-resonant coupling $|2\rangle$-$|4\rangle$.}
\hspace{0in}%
\end{figure}\label{Fig.6}
\end{center}

In Fig.5, we notice that the real part of the relative electric
permittivity $\epsilon_{r}$ maintains negative value by varying p
from 0 to 1. And its value reaches to -2 when p=1. It shows the
intensity of SGC can change the value of Re[$\epsilon_{r}$] but not
change its sign property of plus or minus because of the local field
due to dipole-dipole interaction causing by the high atomic density
here. However, the line shape of Re[$\mu_{r}$] shows different
characteristic when p is in different value range.In the range of
[0, 0.55], Re[$\mu_{r}$] is positive. But it's negative in the range
of [0.55, 1]. And the left-handedness will appear in the range
[0.55, 1] with negative permittivity and permeability
simultaneously. It shows that the magnetic response of the atomic
system is different under the condition of large SGC effect. In the
atomic system, the weak SGC can't cause left-handedness but the
intensifying SGC can do. The refractive index is potted versus p in
Fig.6. The real part of refractive index shows the increasing of SGC
effect can bring about the negative maximum $-2$. However, its
imaginary part illustrates the enhancement of SGC effect can keep
down the absorption, which is certainly favorable for potential
experiments.

From the point of application,the atomic scheme in Fig.1 may be
realized by the Rb atom with 5$S_{1/2}$, 5$P_{3/2}$, 5$D_{3/2}$, and
5$D_{5/2}$ behaving as the$|4\rangle$, $|1\rangle$, $|2\rangle$, and
$|3\rangle$ state labels, respectively. The state 5$D_{3/2}$ is
coupled to the state 5$D_{5/2}$ by a resonant microwave field in
Ref[17]. For another instance,the resonant intermediate level (the
singlet state) between the ground state and the two upper lying
states in the atomic level configuration shown in Fig.1 which is
similar to the level scheme of the experiment mentioned in Ref[20],
and the two highest-lying states are used by a pair of mixed levels
of the singlet and triplet states in sodium dimmer.

\section{Conclusion}

In conclusion, we have investigated the effect of SGC on
left-handedness in the four-level Y-type atom. It was found that the
spontaneously generated coherence plays a significant role in the
realization of left-handedness.When the SGC is weak, the atomic
system doesn't display left-handedness with simultaneous negative
permittivity and permeability. In the case of optimal SGC, the
atomic system has a wide range for simultaneous negative
permittivity and permeability, and the system is left-handedness,and
the absorption is suppressed by the SGC effect when the SGC has a
large intensity. When the probe field near-resonant couples the
atomic system, left-handedness appears until p=0.55, and the maximum
of negative refractive index is -2. We can draw from the above two
cases that the strengthening of SGC produce the left-handedness.And
the stronger the SGC is, the more remarkable the left-handedness
appears facilely.
\section{Acknowledgments}
 The work is supported by the National Natural Science
Foundation of China (Grant Nos.60768001 and 10464002).

\end{document}